\documentclass[11pt]{article}

\usepackage[final]{acl}

\usepackage{times}
\usepackage{latexsym}

\usepackage[T1]{fontenc}

\usepackage[utf8]{inputenc}

\usepackage{microtype}

\usepackage{inconsolata}

\usepackage{graphicx}
\usepackage{tcolorbox}
\usepackage{booktabs}
\usepackage{multirow}
\usepackage{lipsum}

\title{AnyAudio-Judge: A Dynamic Rubric-Based Benchmark and Evaluator for Audio Instruction Following}

\author{
Haitao Li$^{1,2}$,
Tian Tan$^{3}$,
Yuguang Yang$^{4}$,
Shan Yang$^{4}$,
Xie Chen$^{3,2}$\thanks{Corresponding author.} \\
\\
$^1$Zhejiang University \quad $^2$Shanghai Innovation Institute \\
$^3$Shanghai Jiao Tong University \quad $^4$Tencent Hunyuan \\
\texttt{lihaitao@zju.edu.cn, chenxie95@sjtu.edu.cn}
}

\begin{document}
\maketitle
\begin{abstract}
The rapid advancement of instruction-guided audio generation has highlighted the critical need for robust alignment evaluation. Current automated evaluation methods heavily rely on holistic scoring from general-purpose large language models, which struggle to decouple complex instructions, lack interpretability, and fail to capture fine-grained attribute mismatches. To address this, we introduce a novel \textbf{dynamic rubric-based evaluation paradigm} that adaptively decomposes complex audio captions into a variable number of independent, verifiable binary rubric items. To rigorously benchmark this capability, we propose the \textbf{AnyAudio-Judge Bench}, a comprehensive, bilingual benchmark comprising 7,920 meticulously curated samples across four diverse audio domains (speech, sound, music, and mixed), featuring deliberately constructed hard negatives. Furthermore, we construct a large-scale corpus of 105K samples with explicit Chain-of-Thought (CoT) rationales to train our dedicated evaluator, the \textbf{AnyAudio-Judge} model. By employing a training pipeline that combines Supervised Fine-Tuning (SFT) and Group Relative Policy Optimization (GRPO), our model successfully aligns its reasoning paths with the rubric-based scoring mechanism. Extensive experiments demonstrate that AnyAudio-Judge not only significantly enhances zero-shot alignment detection compared to state-of-the-art baselines, but also provides precise and interpretable reward signals that substantially improve instruction alignment in downstream reinforcement learning for audio generation. The benchmark, corpus, and model are available at \url{https://github.com/CuCl-2/AnyAudio-Judge}.
\end{abstract}

\section{Introduction}

Recent advancements in large language models and diffusion models have catalyzed significant progress in instruction-guided audio generation, encompassing zero-shot text-to-speech (InstructTTS) \citep{yang2024instructtts, guo2023prompttts, hu2026qwen3, zhang2025mimo, huang2026moss, li2026restyle}, general sound synthesis \citep{liu2023audioldm, kreuk2022audiogen}, and music generation \citep{copet2023simple}. These foundation models are increasingly capable of following complex, open-ended textual prompts that dictate not only semantic content but also fine-grained acoustic attributes such as emotion, timbre, background environments, and the chronological order of acoustic events. Thus, accurately evaluating the semantic alignment between the generated audio and the complex instructions has emerged as a critical challenge.

Traditionally, evaluating this cross-modal alignment has relied heavily on objective distance metrics (e.g., CLAP similarity \citep{elizalde2023clap}) or human evaluations. While objective embeddings provide a coarse measure of global similarity, they inherently lack the sensitivity to diagnose subtle mismatches in detailed instructions. Conversely, human evaluation is expensive, unscalable, and prone to subjective biases. To mitigate these issues, Large Audio Language Models (LALMs) and general-purpose LLMs (such as Gemini \citep{team2023gemini} or GPT-4 \citep{achiam2023gpt}) have been increasingly adopted as surrogate automated judges. However, most existing instruction-aware evaluators \citep{kuan2026aqascore, huang2025instructttseval, chen2026mint} treat alignment assessment as a monolithic task, employing holistic "yes/no" judgments. This coarse-grained paradigm struggles to decouple highly complex captions, failing to identify exactly \textit{which} specific attributes the generator failed to synthesize. Furthermore, the community still lacks comprehensive, multi-domain benchmarks deliberately designed with hard negative samples to rigorously test the discriminative capabilities of these judge models.

To bridge this gap, we first introduce \textbf{AnyAudio-Judge Bench}, a bilingual, multi-domain benchmark for instruction-audio alignment evaluation. It covers speech, sound, music, and mixed audio, includes both real-world and generated samples, and deliberately constructs hard negatives through instruction swapping and attribute perturbation.

Building on this benchmark, we propose \textbf{AnyAudio-Judge}, an evaluation framework centered on a \textit{dynamic rubric-based evaluation paradigm}. Instead of asking a judge for a single holistic decision, we dynamically derive a variable number ($n$) of independent, verifiable binary rubric items from each instruction and evaluate each item separately. The final score is obtained by aggregating item-level satisfaction probabilities, yielding a more sensitive and interpretable measure of alignment. To support this paradigm beyond black-box prompting, we further construct a 105K-sample \textbf{AnyAudio-Judge Corpus} with hard negatives, generated rubric items, and Chain-of-Thought (CoT) rationales, and train a dedicated evaluator through SFT and GRPO. Finally, we show that the trained judge can serve as a dense reward model for downstream reinforcement learning of instruction-following audio generation.

The main contributions of this work are summarized as follows:
\begin{itemize}
\item We propose \textbf{AnyAudio-Judge Bench}, a bilingual benchmark of \textbf{7,920} curated samples across speech, sound, music, and mixed audio, with hard negatives designed to test fine-grained instruction-audio discrimination.
\item We introduce a \textbf{dynamic rubric-based evaluation paradigm} that decomposes each instruction into a variable number of verifiable binary rubric items and aggregates item-level probabilities into an interpretable alignment score.
\item We construct the \textbf{AnyAudio-Judge Corpus}, a \textbf{105K}-sample training set with hard negatives, rubric items, and CoT rationales, and train \textbf{AnyAudio-Judge} as a dedicated evaluator using SFT and GRPO.
\item We apply AnyAudio-Judge as a reward model for downstream InstructTTS reinforcement learning, demonstrating improved instruction following under judge-guided optimization.
\end{itemize}

\begin{figure*}[t]
    \centering
    \includegraphics[width=1\linewidth]{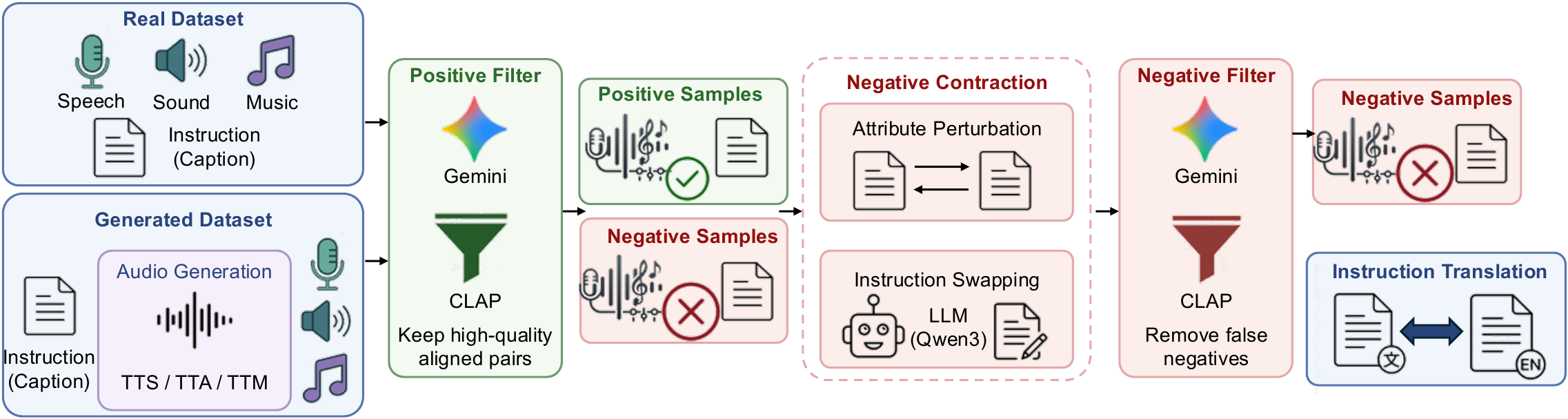}
    \caption{\textbf{AnyAudio-Judge Bench construction pipeline.} We collect real and generated audio across four domains, verify positive pairs, and construct hard negatives via instruction swapping and attribute perturbation. The filtered samples are expanded into symmetric English-Chinese evaluation sets.}
    \label{fig:bench_construction}
\end{figure*}

\section{Related Work}
\textbf{Audio Language Model as a Judge.} Large audio-language models (LALMs) are increasingly used as judges for audio generation, replacing signal-level or embedding-based metrics with natural-language reasoning and feedback. For speech, QualiSpeech \citep{wang2025qualispeech} and TTS-PRISM \citep{wang2026tts} diagnose perceptual defects, while SpeechJudge \citep{zhang2025speechjudge}, GSRM \citep{shen2026gsrm}, and WavReward \citep{ji2025wavreward} study preference judging, reward modeling, and reasoning-enhanced assessment. These works establish LALMs as scalable judges, but mainly emphasize naturalness, expressiveness, or preference consistency rather than fine-grained instruction-audio alignment.

Instruction-guided audio systems expose detailed controls over speaker attributes, emotion, prosody, acoustic events, scene composition, and musical style \citep{yang2024instructtts,guo2023prompttts,hu2026qwen3}. Existing benchmarks test these abilities in narrower settings: InstructTTSEval \citep{huang2025instructttseval} and MINT-Bench \citep{chen2026mint} focus on instruction-following TTS, AQA-Score \citep{kuan2026aqascore} targets text-to-audio alignment, and CMI-Reward \citep{ma2026cmi} evaluates music reward models. AnyAudio-Judge instead studies unified judging across speech, sound, music, and mixed audio, with hard negatives that expose subtle compositional failures.

\textbf{Rubric-based evaluation and rewards.} Rubric-based evaluation decomposes holistic judgment into explicit criteria, improving transparency over a single scalar score. In language tasks, analytic rubric frameworks and LLM-based evaluators improve interpretability and calibration \citep{pathak2025rubric,ye2023flask,hashemi2024llm}, while AutoRubric learns reusable reward criteria from preferences \citep{xie2025auto}. AutoRubric-R1V derives problem-specific rubrics from successful multimodal reasoning trajectories \citep{jia2025autorubric}. Yet these methods are mostly text-centered, fixed or reusable in scope, or aimed at reasoning supervision rather than audio-instruction verification. AnyAudio-Judge brings this paradigm to instruction-following audio generation through dynamic, instance-specific binary rubric items.

\section{AnyAudio-Judge Bench}
\label{sec:anyaudio_judge_bench}

Most current evaluation paradigms for instruction-following audio models rely on general-purpose large language models, such as Gemini \citep{team2023gemini}, as surrogate judges. However, the judgments produced by these models often fail to align with human perception. More importantly, the community still lacks dedicated benchmarks to rigorously evaluate how well these judge models actually discriminate between aligned and misaligned audio. To address this gap, we introduce the \textbf{AnyAudio-Judge Bench}. Figure~\ref{fig:bench_construction} summarizes its construction pipeline.

\begin{figure*}[t]
    \centering
    \includegraphics[width=1\linewidth]{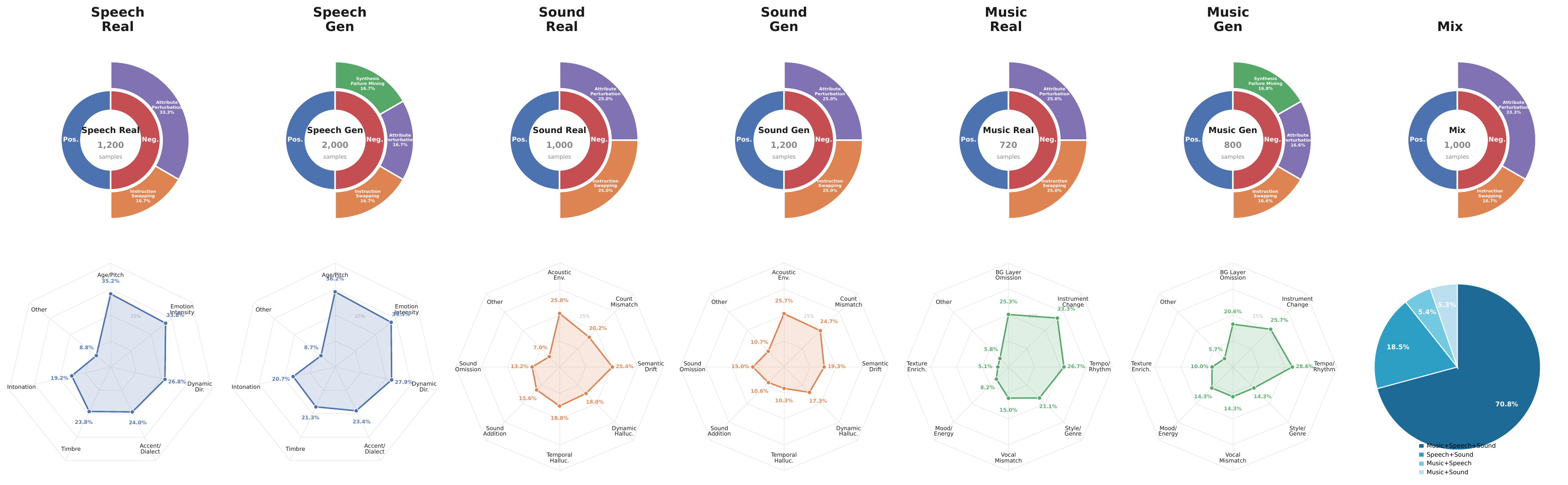}
    \caption{\textbf{Statistics of AnyAudio-Judge Bench.} Upper: subset composition; lower: negative-type distribution per subset and domain composition of \textsc{Mix}.}
    \label{fig:benchmark_stats}
\end{figure*}

Our benchmark covers a wide range of audio modalities, including speech, sound, music, and mixed audio, and incorporates both real-world recordings and synthesized outputs. To evaluate cross-lingual generalization, we built fully symmetric bilingual evaluation sets in English and Chinese. These are divided into seven distinct subsets: \textsc{Speech-Real}, \textsc{Speech-Gen}, \textsc{Sound-Real}, \textsc{Sound-Gen}, \textsc{Music-Real}, \textsc{Music-Gen}, and \textsc{Mix}. Because high-fidelity mixed-audio synthesis remains challenging, the \textsc{Mix} subset consists entirely of real-world samples. Across all subsets, we maintain a strict 1:1 ratio of positive to negative samples to ensure a balanced evaluation. Figure~\ref{fig:benchmark_stats} details the dataset statistics.

\noindent\textbf{General Negative Construction Paradigm.} Evaluating a judge model differs fundamentally from evaluating an audio generator: it requires high-quality negative samples to test whether the judge can successfully detect when audio and text do not match. After verifying positive pairs across all domains, we construct challenging negative samples using two main strategies. The first is \textbf{Instruction Swapping}, where we interchange instructions between different samples to create clear semantic mismatches. The second is \textbf{Attribute Perturbation}, which uses an LLM (e.g., Qwen3-30B-A3B-Instruct-2507 \citep{yang2025qwen3}) to alter specific details in the original captions, simulating fine-grained alignment failures. Detailed prompts used for negative sample construction and dataset filtering are provided in Appendix~\ref{sec:prompt_templates}.

\noindent\textbf{Speech Subsets.} For the \textsc{Speech-Real} subset, we use audio and descriptions from the InstructTTSEval dataset \citep{huang2025instructttseval} as our initial positive samples. We then generate negative samples using the two methods described above. To ensure our ground-truth labels are reliable, we use Gemini to filter out ambiguous cases, specifically discarding swapped pairs where the new instruction might still plausibly describe the audio.

For \textsc{Speech-Gen}, we prompt several state-of-the-art generators, including Qwen3TTS-12Hz-1.7B-VD \citep{hu2026qwen3}, MOSS-VoiceGenerator \citep{huang2026moss}, and MiMo-Audio-7B-Instruct \citep{zhang2025mimo}, using instructions from the real subset. Since synthesized speech is not always perfect, we use Gemini to evaluate the outputs. We specifically mine genuine synthesis failures to serve as hard negatives by running a dual-pass Gemini evaluation, keeping only the failures that both passes agree on (resulting in an approximate 50\% overlap). Successful generations become our positive samples. We then expand the negative set for these positive samples using our swapping and perturbation methods, such as simulating inaccurate dialects or missing emotional tone.

\noindent\textbf{Sound Subsets.} The \textsc{Sound-Real} subset draws from the Clotho v2 test set \citep{drossos2020clotho}. We filter this data using CLAP \citep{elizalde2023clap}, keeping only pairs with a similarity score $\ge 0.6$ as our base positives, and then construct negative samples through swapping and perturbation. When perturbing attributes, we specifically instruct the LLM to mimic common text-to-audio failure modes, such as dropping secondary sounds or disrupting the chronological order of acoustic events. To ensure the swapped negative samples are genuinely mismatched, we apply a reverse CLAP filter to remove any pairs that still score $> 0.5$. 

For \textsc{Sound-Gen}, we synthesize clips using AudioGen \citep{kreuk2022audiogen}, AudioLDM2 \citep{liu2024audioldm}, and Stable Audio \citep{evans2025stable}. Because generation quality can fluctuate, we again use CLAP to confirm which outputs are semantically aligned and treat those as positive samples. We then build the corresponding negative samples using our two standard methods.

\noindent\textbf{Music Subsets.} The \textsc{Music-Real} subset is based on the Song Describer Dataset \citep{manco2023song}. We use Gemini with tailored prompts to filter the data, retaining only highly accurate captions as our positive samples. After generating negatives via swapping and perturbation, we run a second Gemini verification to discard any newly formed pairs that do not present a clear semantic mismatch.

For \textsc{Music-Gen}, we synthesize tracks using MusicGen \citep{copet2023simple}, ACE-Step \citep{gong2025ace}, and Stable Audio \citep{evans2025stable}. Gemini evaluates these outputs to form initial sets of positive and negative samples based on how well they follow the instructions. We then expand the negative set using swapping and perturbation, followed by a final Gemini check to verify the accuracy of all constructed negatives.

\noindent\textbf{Mixed Audio Subsets.} The \textsc{Mix} subset consists of cinematic audio tracks segmented into roughly one-minute clips. Because these tracks do not come with instructions, we use Gemini to generate comprehensive captions for them, which serve as our positive samples. Finally, we construct challenging semantic negatives using the same swapping and perturbation pipelines applied to the other modalities.

\noindent\textbf{Symmetric Bilingual Benchmark.} We build a fully symmetric English-Chinese benchmark to test cross-lingual robustness. For every sample, we translate the textual instruction in both directions, yielding parallel English and Chinese evaluation sets. Only the prompts are translated; all spoken content is left unchanged.

\begin{figure*}[t]
    \centering
    \includegraphics[width=1.0\linewidth]{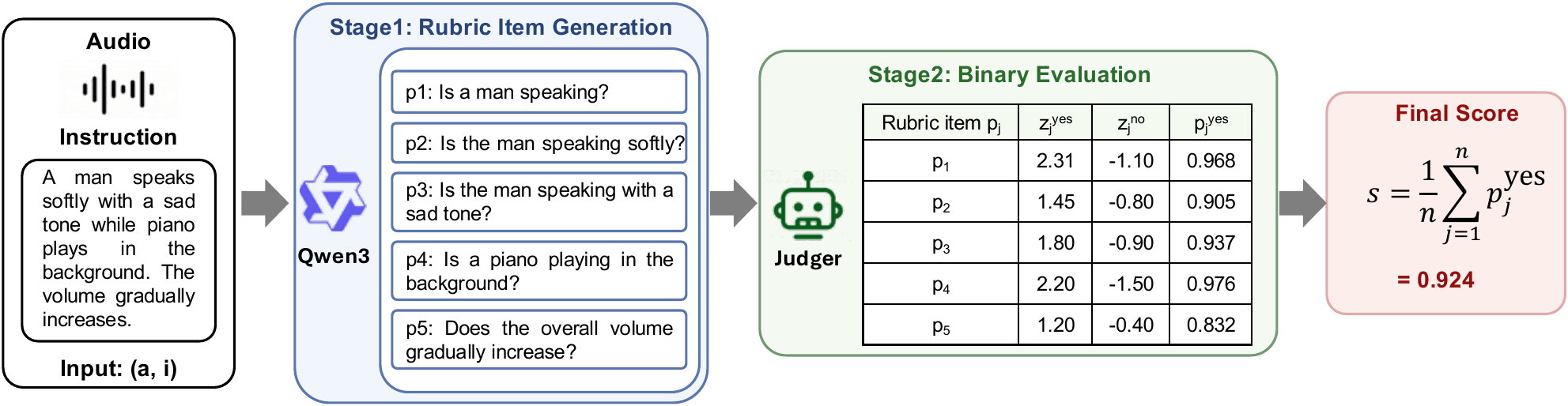}
    \caption{\textbf{Overview of AnyAudio-Judge.} Given an audio-instruction pair, the instruction is first decomposed into dynamic rubric items. The judge then evaluates each item with a yes/no probability derived from the corresponding logits, and aggregates the item-level probabilities into an interpretable alignment score.}
    \label{fig:judge_overview}
\end{figure*}

\section{Method}
\label{sec:method}
\subsection{Dynamic Rubric-based Evaluation Paradigm}

To address the limitations of traditional holistic judgment methods, we introduce a dynamic rubric-based evaluation paradigm that decomposes complex audio captions into a set of binary True/False rubric items, as illustrated in Figure~\ref{fig:judge_overview}. Given an audio-instruction pair $(a, i)$, where $a$ is the audio and $i$ is the instruction, our paradigm first decomposes $i$ into $n$ atomic rubric items $\{p_1, p_2, \dots, p_n\}$. The decomposition is performed by an LLM (Qwen3-30B-A3B-Instruct-2507) using a structured prompt that instructs it to break down the instruction into independent, directly verifiable statements; the decomposition and question-filtering prompts are provided in Appendix~\ref{sec:prompt_decompose}. Each rubric item is constructed so that a "yes" answer indicates alignment between the audio and the corresponding aspect of the instruction, making the final score semantically meaningful.

For each rubric item $p_j$, the judge model evaluates whether the audio satisfies the corresponding criterion by comparing the logits of the two candidate answers, ``yes'' and ``no''. Let $z_j^{\mathrm{yes}}$ and $z_j^{\mathrm{no}}$ denote the logits assigned to these two answers for the $j$-th rubric item. We normalize them with a two-way softmax and obtain the soft satisfaction probability:

\[
p_j^{\mathrm{yes}} =
\frac{\exp(z_j^{\mathrm{yes}})}
{\exp(z_j^{\mathrm{yes}}) + \exp(z_j^{\mathrm{no}})}
\]

The overall alignment score $s$ is then computed as the average yes probability over all rubric items:

\[
s = \frac{1}{n}\sum_{j=1}^{n} p_j^{\mathrm{yes}}
\]

This rubric-based formulation improves interpretability by exposing item-level alignment failures, and it also enhances zero-shot evaluation by converting holistic audio-text matching into a set of binary verification problems. In practice, we implement this paradigm using a two-stage process: (1) rubric item generation via LLM, and (2) binary evaluation by the judge model.

\begin{table*}[t]
\centering
\resizebox{\textwidth}{!}{
\begin{tabular}{ll *{8}{c}}
\toprule
\multirow{2}{*}{Model} & \multirow{2}{*}{Prompt} & \multicolumn{8}{c}{\textbf{Chinese (zh)}} \\
\cmidrule(lr){3-10}
& & \textsc{Speech-Real} & \textsc{Speech-Gen} & \textsc{Sound-Real} & \textsc{Sound-Gen} & \textsc{Music-Real} & \textsc{Music-Gen} & \textsc{Mix} & \textbf{Avg} \\
\midrule
\multirow{2}{*}{Audio-Flamingo3} 
& holistic & 50.25 & 50.35 & 65.90 & 67.66 & 59.97 & 59.52 & 52.11 & 57.97 \\
& dynamic rubric & 60.25 & 58.90 & 65.17 & 68.92 & 70.28 & 67.37 & 56.50 & 63.91 \\
\midrule 
\multirow{2}{*}{MiDashengLM} 
& holistic & 51.58 & 52.50 & 53.40 & 49.50 & 52.22 & 48.88 & 47.20 & 50.75 \\
& dynamic rubric & 66.94 & 65.50 & 67.90 & 75.83 & 71.25 & 71.13 & 59.90 & 68.35 \\
\midrule
\multirow{2}{*}{Kimi-Audio-7B-Instruct} 
& holistic & 51.00 & 50.20 & 52.50 & 50.25 & 52.08 & 51.50 & 49.00 & 50.93 \\
& dynamic rubric & 81.17 & 70.10 & 66.80 & 75.17 & 69.72 & 68.00 & 64.90 & 70.84 \\
\midrule
\multirow{2}{*}{Qwen2.5-Omni-7B} 
& holistic & 50.25 & 50.20 & 50.40 & 51.25 & 56.25 & 53.88 & 50.10 & 51.76 \\
& dynamic rubric & 78.17 & 72.10 & 68.00 & 75.58 & 74.72 & 71.63 & 63.30 & 71.93 \\
\midrule
\multirow{2}{*}{Qwen3-Omni-30B-A3B-Instruct} 
& holistic & 67.33 & 58.10 & 54.10 & 57.92 & 64.58 & 66.75 & 52.90 & 60.24 \\
& dynamic rubric & 90.75 & 78.65 & 71.60 & \underline{77.33} & 76.25 & 75.88 & 67.30 & 76.82 \\
\midrule
\multirow{2}{*}{Qwen3-Omni-30B-A3B-Captioner} 
& holistic & 75.42 & 64.75 & 59.20 & 64.00 & 68.61 & 70.00 & 55.30 & 65.33 \\
& dynamic rubric & 90.17 & 79.50 & 70.80 & 76.42 & 74.86 & 75.75 & 69.10 & 76.66 \\
\midrule
\multirow{2}{*}{Gemini-2.5-Pro} 
& holistic & \underline{92.33} & \textbf{81.40} & 69.42 & 76.40 & \underline{85.63} & \textbf{81.47} & 73.40 & \underline{80.01} \\
& dynamic rubric & 90.67 & 80.00 & \underline{72.00} & 77.00 & 79.67 & 73.33 & \underline{75.50} & 78.31 \\
\midrule
\textbf{AnyAudio-Judge} & dynamic rubric & \textbf{93.33} & \underline{80.15} & \textbf{77.90} & \textbf{82.50} & \textbf{92.22} & \underline{80.13} & \textbf{90.60} & \textbf{85.26} \\
\bottomrule
\end{tabular}
}
\caption{Results on the AnyAudio-Judge Benchmark (Chinese Subset).}
\label{tab:results_zh}
\end{table*}

\begin{table*}[t]
\centering
\resizebox{\textwidth}{!}{
\begin{tabular}{ll *{8}{c}}
\toprule
\multirow{2}{*}{Model} & \multirow{2}{*}{Prompt} & \multicolumn{8}{c}{\textbf{English (en)}} \\
\cmidrule(lr){3-10}
& & \textsc{Speech-Real} & \textsc{Speech-Gen} & \textsc{Sound-Real} & \textsc{Sound-Gen} & \textsc{Music-Real} & \textsc{Music-Gen} & \textsc{Mix} & \textbf{Avg} \\
\midrule
\multirow{2}{*}{Audio-Flamingo3} 
& holistic & 48.87 & 51.26 & 63.78 & 67.58 & 69.32 & 64.78 & 51.65 & 59.61 \\
& dynamic rubric & 59.00 & 58.40 & 67.60 & 68.42 & 70.97 & 67.75 & 57.20 & 64.19 \\
\midrule 
\multirow{2}{*}{MiDashengLM} 
& holistic & 48.58 & 48.95 & 50.20 & 51.33 & 50.56 & 50.75 & 47.60 & 49.71 \\
& dynamic rubric & 64.92 & 63.60 & 69.00 & 74.75 & 72.78 & 70.63 & 59.90 & 67.94 \\
\midrule
\multirow{2}{*}{Kimi-Audio-7B-Instruct} 
& holistic & 50.25 & 50.15 & 51.30 & 49.58 & 50.83 & 50.38 & 48.60 & 50.16 \\
& dynamic rubric & 80.08 & 69.15 & 68.00 & 74.00 & 70.42 & 68.25 & 65.80 & 70.81 \\
\midrule
\multirow{2}{*}{Qwen2.5-Omni-7B} 
& holistic & 50.25 & 50.20 & 50.40 & 51.25 & 56.25 & 53.88 & 50.10 & 51.76 \\
& dynamic rubric & 77.25 & 69.90 & 67.40 & 76.25 & 77.08 & 71.37 & 66.40 & 72.24 \\
\midrule
\multirow{2}{*}{Qwen3-Omni-30B-A3B-Instruct} 
& holistic & 65.42 & 58.45 & 54.10 & 58.75 & 68.61 & 70.25 & 52.50 & 61.15 \\
& dynamic rubric & 88.92 & 77.85 & \underline{73.30} & \underline{78.92} & 78.89 & 77.38 & 66.10 & 77.34 \\
\midrule
\multirow{2}{*}{Qwen3-Omni-30B-A3B-Captioner} 
& holistic & 70.46 & 61.70 & 60.40 & 66.58 & 67.92 & 70.75 & 51.90 & 64.24 \\
& dynamic rubric & 88.50 & 77.35 & 72.10 & 78.42 & 78.33 & 76.00 & 66.70 & 76.77 \\
\midrule
\multirow{2}{*}{Gemini-2.5-Pro} 
& holistic & \underline{91.22} & \textbf{79.20} & 70.23 & 72.60 & \underline{83.25} & \underline{78.87} & 69.80 & \underline{77.72} \\
& dynamic rubric & 89.00 & 77.67 & 71.67 & 76.00 & 79.33 & 76.19 & \underline{71.00} & 77.27 \\
\midrule
\textbf{AnyAudio-Judge} & dynamic rubric & \textbf{91.42} & \underline{78.85} & \textbf{78.70} & \textbf{84.67} & \textbf{91.25} & \textbf{79.87} & \textbf{86.40} & \textbf{84.45} \\
\bottomrule
\end{tabular}
}
\caption{Results on the AnyAudio-Judge Benchmark (English Subset).}
\label{tab:results_en}
\end{table*}

\subsection{AnyAudio-Judge Corpus}

The AnyAudio-Judge Corpus is a 105K-sample training set constructed to support our dynamic rubric-based evaluation paradigm beyond benchmark-only evaluation. It is built from data sources disjoint from the benchmark, making the benchmark an out-of-distribution evaluation set relative to the training corpus, and provides two annotation layers for each sample: per-rubric binary labels and explicit Chain-of-Thought (CoT) rationales.

\textbf{Data Sources and Negative Construction.} To cover a broader real-world distribution, we collect positive audio-caption pairs from four domains: internal speech assets with high-quality annotations (Speech), AudioCaps \citep{kim2019audiocaps} (Sound), the MusicBench training split \citep{melechovsky2024mustango} (Music), and new movie audio clips with Gemini-verified captions (Mixed). From these base pairs, we construct hard negatives using the same pipeline as the benchmark, including instruction swapping and attribute perturbation. The resulting corpus contains both coarse mismatches and subtle semantic drifts.

\textbf{Annotation Pipeline.} The annotation pipeline contains three steps. First, we decompose the original caption into $n$ binary rubric items via LLM prompting. Second, we pair each audio clip with either its positive caption or a constructed negative caption. Finally, we generate per-rubric yes/no labels and CoT rationales by comparing the original true caption with the target caption.

\textbf{Per-rubric Judgment Generation.} For positive samples, every rubric item is labeled as ``yes,'' with a rationale stating that the referenced feature appears in the audio. For negative samples, however, the per-rubric labels are not uniformly ``no.'' We derive them through a text-only LLM using Qwen3-30B-A3B-Instruct-2507: given the original caption as the true audio description and the modified caption as the negative hypothesis, the LLM determines the answer for each rubric item and produces a CoT rationale for the comparison. This enables fine-grained contradiction supervision, since a negative caption may still preserve some correct attributes while perturbing others.

\textbf{Corpus Characteristics.} The resulting SFT corpus contains explicit reasoning chains and balanced positive-negative pairs. In total, it includes 30K Speech, 30K Sound, 30K Music, and 15K Mixed audio samples, with a strict 1:1 positive-to-negative ratio within each modality.

\subsection{AnyAudio-Judge Model}

Leveraging the AnyAudio-Judge Corpus, we train AnyAudio-Judge with a two-stage pipeline: Supervised Fine-Tuning (SFT) followed by Group Relative Policy Optimization (GRPO) \citep{shao2024deepseekmath}. The goal is to obtain a judge that not only predicts rubric-level yes/no labels, but also produces concise evidence for each decision.

\textbf{Supervised Fine-Tuning (SFT):} We initialize from Qwen3-Omni-30B-A3B-Captioner and fine-tune on the corpus to establish the basic rubric-following behavior. Given an audio-instruction pair and its decomposed rubric items, the model learns to output binary judgments together with Chain-of-Thought (CoT) rationales.

\textbf{Group Relative Policy Optimization (GRPO):} To further improve accuracy on ambiguous cases, we apply GRPO after SFT. We first perform four rollouts on the AnyAudio-Judge Corpus and remove samples that are consistently answered correctly, leaving 8,454 harder samples for optimization. The reward is a weighted sum of three terms:
\begin{itemize}
    \item \textbf{Format consistency (0.1)} checks whether the output is a valid JSON and each record contains the required fields, including the rubric ID, binary answer, and supporting evidence.
    \item \textbf{Global accuracy (0.2)} evaluates the overall matched/mismatched decision induced by the rubric answers: all ``yes'' answers imply a matched sample, while any ``no'' answer implies a mismatch.
    \item \textbf{Balanced rubric accuracy (0.7)} measures fine-grained rubric-level correctness by averaging the accuracy on gold ``yes'' and gold ``no'' rubric items, reducing the incentive to over-predict either class.
\end{itemize}
Both stages are trained on 16 H20 GPUs (96GB). SFT uses full-parameter fine-tuning for one epoch, with a per-device batch size of 4, gradient accumulation of 1, and a learning rate of $1\times10^{-5}$. GRPO uses LoRA with rank 16 and alpha 32 for one epoch, with 8 generations per prompt, a per-device batch size of 8, gradient accumulation of 1, and a learning rate of $5\times10^{-6}$.
The trained model outputs a JSON array for each audio-instruction pair, with each record containing the rubric ID, binary answer, and supporting evidence.

\begin{table}[t]
\centering
\resizebox{0.9\columnwidth}{!}{
\begin{tabular}{l ccc}
\toprule
\multirow{2}{*}{\textbf{Method / Model}} & \multicolumn{3}{c}{PAM} \\
\cmidrule(lr){2-4}
& LCC $\uparrow$ & SRCC $\uparrow$ & KTAU $\uparrow$ \\
\midrule
CLAPScore & 0.472 & 0.477 & 0.337 \\
\midrule
\multicolumn{4}{l}{\textit{AQAScore}} \\
\quad Qwen2.5-Omni-3B & 0.540 & 0.560 & 0.410 \\
\quad Qwen2.5-Omni-7B & 0.518 & \underline{0.589} & \underline{0.429} \\
\quad AF3             & 0.496 & 0.538 & 0.383 \\
\quad AF3-Think       & \underline{0.582} & 0.587 & 0.419 \\
\quad AF3-Chat        & 0.381 & 0.435 & 0.337 \\
\midrule
\textbf{AnyAudio-Judge} & \textbf{0.614} & \textbf{0.601} & \textbf{0.435} \\
\bottomrule
\end{tabular}
}
\caption{Results on the PAM dataset.}
\label{tab:relate_pam_comparison}
\end{table}

\section{Evaluation of AnyAudio-Judge}

We evaluate AnyAudio-Judge from three complementary perspectives: alignment detection on our constructed AnyAudio-Judge Bench, generalization to an external benchmark, and ablations over the evaluation paradigm and training stages.

\subsection{Experimental Setup and Metrics}

On AnyAudio-Judge Bench, each method is evaluated separately on the Chinese and English subsets, with classification accuracy (ACC) as the metric. For LALM baselines \citep{ghosh2026audio,dinkel2025midashenglm,ding2025kimi,xu2025qwen3}, we test both holistic prompting and dynamic rubric prompting: the former asks for one match/mismatch decision, while the latter decomposes the instruction into verifiable binary items and aggregates item-level judgments. The two judge prompt templates are provided in Appendix~\ref{sec:prompt_judge}.
To further verify generalization, we evaluate on the external PAM dataset \citep{deshmukh2024pam} and report Pearson linear correlation (LCC), Spearman rank correlation (SRCC), and Kendall's tau (KTAU).

\subsection{Results on the AnyAudio-Judge Benchmark}

The benchmark results in Tables~\ref{tab:results_zh} and~\ref{tab:results_en} show a consistent pattern across Chinese and English: dynamic rubric prompting substantially improves most LALM baselines over holistic judgment. This suggests that explicit, item-level checks are more effective than a single global decision for fine-grained instruction-audio alignment.

However, prompt-only rubric evaluation remains limited by the reasoning and audio grounding ability of the base LALM, motivating a dedicated evaluator aligned with the rubric paradigm.

With dedicated training, AnyAudio-Judge achieves the best average accuracy on both subsets, reaching 85.26 on Chinese and 84.45 on English. The gains are especially clear on fine-grained and mixed-domain subsets such as \textsc{Sound-Gen}, \textsc{Music-Real}, and \textsc{Mix}, indicating that rubric-level supervision and rationales improve performance.

\subsection{Generalization on External Benchmarks}

On PAM, AnyAudio-Judge obtains the strongest correlation with human preferences across all three metrics, outperforming both CLAPScore \citep{elizalde2023clap} and AQAScore \citep{kuan2026aqascore} variants (Table~\ref{tab:relate_pam_comparison}). The result indicates that explicit rubric supervision transfers beyond our benchmark construction and provides value beyond global embedding similarity.

\subsection{Ablation Studies}

The ablation in Table~\ref{tab:ablation} separates the contribution of the evaluation paradigm from the training pipeline. Moving from holistic judgment to dynamic rubric evaluation already yields a large improvement on both languages, confirming the importance of explicit decomposition.
SFT further teaches the model to follow the rubric format and output item-level judgments. GRPO provides an additional gain by focusing optimization on harder samples that remain ambiguous after supervised fine-tuning.

\begin{table}[t]
\centering
\small
\resizebox{\columnwidth}{!}{
\begin{tabular}{lcc}
\toprule
\textbf{Method} & \textbf{Chinese ACC} & \textbf{English ACC} \\
\midrule
Holistic judgment & 65.33 & 64.24 \\
Dynamic rubric & 76.66 & 76.77 \\
+ SFT & 84.02 & 83.78 \\
+ SFT + GRPO & \textbf{85.26} & \textbf{84.45} \\
\bottomrule
\end{tabular}
}
\caption{Ablation study on evaluation and training strategies.}
\label{tab:ablation}
\end{table}

\subsection{Analysis of Dynamic Rubric Items}

\begin{figure}[t]
    \centering
    \includegraphics[width=0.8\linewidth]{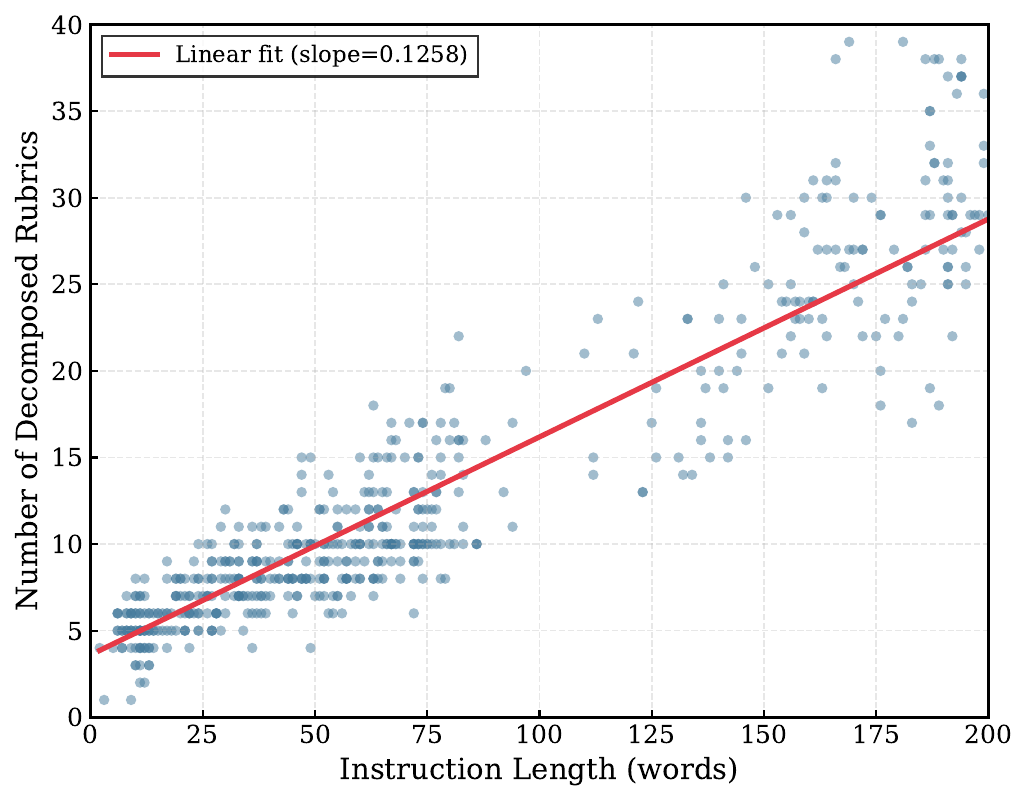}
    \caption{Dynamic granularity analysis: the number of decomposed binary rubric items ($n$) adapts to instruction complexity.}
    \label{fig:dynamic_granularity}
\end{figure}

The number of generated rubric items adapts to instruction complexity (Figure~\ref{fig:dynamic_granularity}). Simple instructions require only a few checks, while detailed captions are decomposed into more items, allowing the evaluator to allocate granularity where the instruction contains more verifiable constraints.

\section{Applications of AnyAudio-Judge}

Having verified the evaluation capability of AnyAudio-Judge, we further use it as a practical tool in two scenarios: assessing the quality of instruction-following generators and serving as a reward model for reinforcement learning.

\subsection{Evaluating InstructTTS Models with AnyAudio-Judge}

\begin{table}[t]
\centering
\small
\resizebox{0.9\columnwidth}{!}{
\begin{tabular}{lc}
\toprule
\textbf{Generator Model} & \textbf{Judge Score} \\
\midrule
\multicolumn{2}{l}{\textit{Commercial Systems}} \\
Gemini 2.5-Pro & 87.5 \\
\midrule
\multicolumn{2}{l}{\textit{Open-source Systems}} \\
Qwen3-TTS-12Hz-1.7B-VD & 84.8 \\
MOSS-VoiceGenerator & 80.6 \\
MiMo-Audio-7B-Instruct & 81.1 \\
\bottomrule
\end{tabular}
}
\caption{Evaluation of state-of-the-art InstructTTS models using AnyAudio-Judge.}
\label{tab:tts_eval}
\end{table}

We first apply the judge to representative InstructTTS systems, including Qwen3-TTS-12Hz-1.7B-VD \citep{hu2026qwen3}, MOSS-VoiceGenerator \citep{huang2026moss}, and MiMo-Audio-7B-Instruct \citep{zhang2025mimo}. As shown in Table~\ref{tab:tts_eval}, Gemini 2.5-Pro receives the highest score, while Qwen3-TTS-12Hz-1.7B-VD is the strongest open-source model. Because the score is tied to item-level instruction satisfaction, it remains sensitive to partial failures in speaker, prosody and style.

\subsection{Reward Model for InstructTTS Reinforcement Learning}
\begin{figure}[t]
    \centering
    \includegraphics[width=\linewidth]{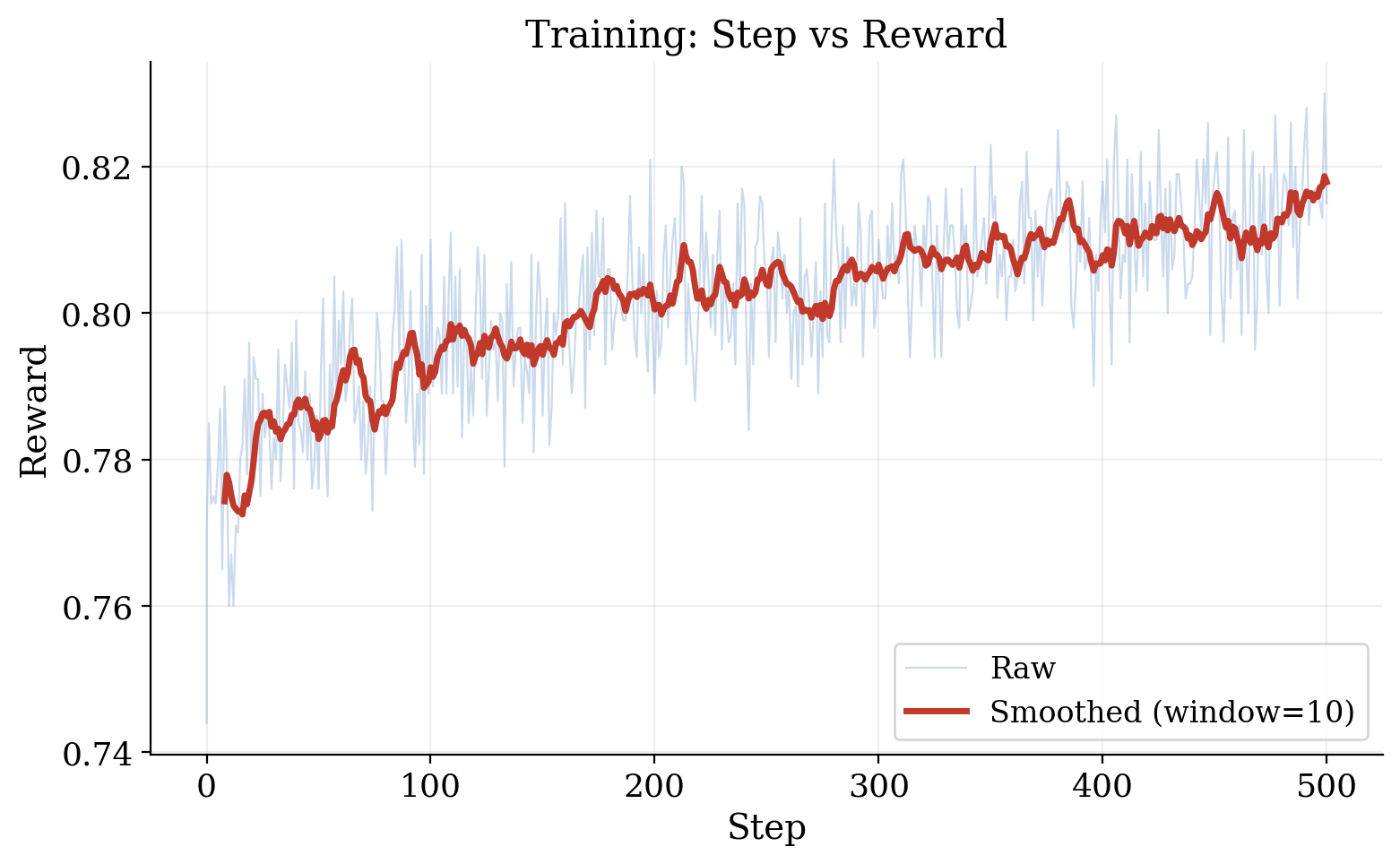}
    \caption{Reward trajectory during InstructTTS reinforcement learning with AnyAudio-Judge as reward.}
    \label{fig:reward_training_curve}
\end{figure}

For reward modeling, we use DiTAR \citep{jia2025ditar} as the base model and optimize it with GRPO using AnyAudio-Judge scores as rewards. For each generated sample, the judge estimates item-level satisfaction and aggregates the probabilities into a scalar reward, which is denser than a binary preference label and more interpretable than an embedding score.

\begin{figure}[t]
    \centering
    \includegraphics[width=\linewidth]{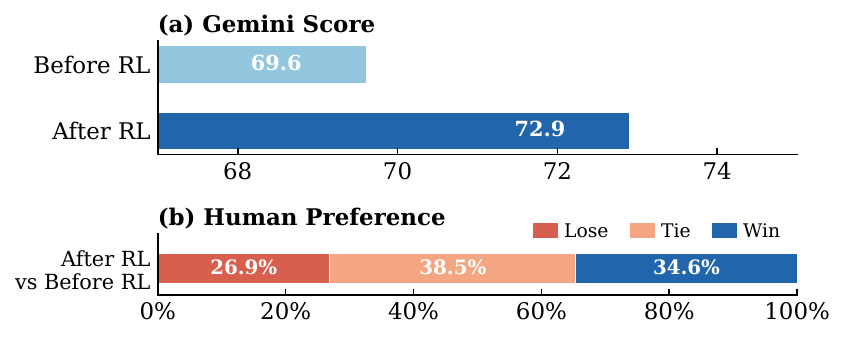}
    \caption{Human preference and Gemini score comparison for InstructTTS models fine-tuned with different reward signals.}
    \label{fig:human_win_rate}
\end{figure}

During optimization, the reward rises steadily, indicating that the model learns to satisfy more rubric items over training steps (Figure~\ref{fig:reward_training_curve}).
We further evaluate the fine-tuned models on InstructTTSEval \citep{huang2025instructttseval} using both Gemini-based scores and human preference judgments. The AnyAudio-Judge-optimized model outperforms the base model (Figure~\ref{fig:human_win_rate}), suggesting that the learned reward provides a robust optimization signal.

\section{Conclusion}
We presented AnyAudio-Judge, a dynamic rubric-based framework for evaluating instruction-audio alignment. By decomposing complex instructions into verifiable binary rubric items, AnyAudio-Judge provides a more fine-grained and interpretable alternative to holistic audio-text scoring. Together with AnyAudio-Judge Bench and the 105K-sample AnyAudio-Judge Corpus, our trained evaluator improves alignment detection, generalizes to external preference evaluation, and serves as an effective reward model for InstructTTS reinforcement learning.

\section*{Limitations}
AnyAudio-Judge depends on the quality of rubric decomposition. While dynamic rubric items improve interpretability, imperfect decomposition may miss implicit constraints or split a single attribute too finely. The judge may therefore perform worse when rubric generation is of insufficient quality, and rubric generation also introduces additional inference time.

\section*{Acknowledgments}
This work was supported by National Natural Science Foundation of China  (No. U23B2018), the Science and Technology Innovation (STI) 2030-Major Project (2022ZD0208700), and Yangtze River Delta Science and Technology Innovation Community Joint Research Project (2024CSJGG1100).

\bibliography{custom}

\appendix
\section{Prompt Templates for Benchmark Construction}
\label{sec:prompt_templates}

In this section, we present the exact prompts utilized in constructing the AnyAudio-Judge Bench. Tables~\ref{tab:prompt_attr_speech}--\ref{tab:prompt_attr_music} show the attribute perturbation prompts used to construct hard negatives for the speech, sound, and music subsets, respectively. Tables~\ref{tab:prompt_filter_speech} and~\ref{tab:prompt_filter_music} provide the filtering prompts used to verify speech and music data quality during benchmark construction.

\begin{table*}[t]
\centering
\small
\begin{tcolorbox}[colback=gray!10, colframe=black, boxrule=0.5pt, arc=3pt, width=\textwidth]
\textbf{System Prompt:} \\
You are an expert in audio description modification. Given a description that matches an audio clip, please perform \textbf{targeted, fine-grained} modifications to make the modified description no longer match the original audio, thereby constructing hard-to-distinguish negative samples.

\vspace{0.5em}
\noindent\textbf{Modification Strategies} (Choose 1-2 types that best suit the current description):
\begin{enumerate}
    \setlength\itemsep{0em}
    \item \textbf{Reversal of Dynamic Changes:} If the description involves a gradual process (e.g., ``from slow to fast,'' ``from excited to calm,'' ``gradually increasing''), change it to the opposite direction or to ``stable with no obvious changes.''
    \item \textbf{Adjustment of Emotion Intensity:} If the described emotion is relatively calm, change it to strong emotions like ``passionate,'' ``indignant,'' or ``furious''; if the described emotion is already strong, change it to ``calm and restrained'' or ``deep and introverted.''
    \item \textbf{Reversal of Terminal Intonation:} If the description requires ``rising intonation at the end of the sentence,'' change it to ``falling intonation''; and vice versa. If there is no intonation requirement, add an intonation requirement that contradicts a flat, steady reading.
    \item \textbf{Accent/Dialect Substitution:} Change a requirement for standard Mandarin to ``heavy Cantonese accent'' or ``obvious dialect features''; or change a dialect requirement to standard Mandarin.
    \item \textbf{Timbre/Texture Substitution:} Change ``bright and clear'' to ``hoarse,'' ``grainy,'' or ``deep and gloomy''; or vice versa.
    \item \textbf{Age/Pitch Level Substitution:} Change ``middle-aged male/female'' to ``elderly'' or ``young''; change ``low pitch'' to ``high-pitched and bright''; or vice versa.
\end{enumerate}

\noindent\textbf{Modification Principles:}
\begin{itemize}
    \setlength\itemsep{0em}
    \item Modify only 1-2 places, keeping the rest unchanged to ensure the overall description remains fluent and reasonable.
    \item Modifications must precisely target specific, perceivable acoustic attributes without changing the basic scene type.
    \item \textbf{Internal Consistency:} After modifying a certain attribute, synchronously check and adjust other related details in the description (e.g., update pronouns and timbre wording when changing gender; update accompanying descriptions like speaking rate and volume when changing emotion intensity), ensuring the modified overall description is not self-contradictory.
    \item Output the modified description directly without adding any explanations.
\end{itemize}

\vspace{0.5em}
\noindent\textbf{Original description:} \{caption\} \\
\textbf{Modified description:}
\end{tcolorbox}
\caption{Prompt of Attribute Perturbation for the Speech subset, simulating text-to-speech (TTS) generation failures.}
\label{tab:prompt_attr_speech}
\end{table*}

\begin{table*}[t]
\centering
\small
\begin{tcolorbox}[colback=gray!10, colframe=black, boxrule=0.5pt, arc=3pt, width=\textwidth]
\textbf{System Prompt:} \\
You are simulating the failure modes of a text-to-audio (TTA) generation model. Given a caption that accurately describes an audio clip, rewrite it to describe what a flawed TTA model might actually generate — audio that diverges from the original in ways characteristic of real TTA errors. The result is a ``negative caption'' describing the TTA model's erroneous output.

\vspace{0.5em}
\noindent\textbf{Common TTA failure modes to simulate} (apply 2–3 that are most applicable):
\begin{enumerate}
    \setlength\itemsep{0em}
    \item \textbf{Secondary sound omission:} Simulate this by ADDING a plausible background or secondary sound to the caption that the real audio likely does not contain. The caption now expects a sound the audio is missing.
    \item \textbf{Spurious sound addition:} Simulate this by OMITTING a sound from the caption that is present in the real audio. Do not say ``no seagulls''; simply describe the scene without them.
    \item \textbf{Temporal sequence enrichment:} TTA often outputs a flat blend of sounds, losing order. Simulate this by describing a clear, vivid temporal sequence that the flat audio does not actually contain.
    \item \textbf{Dynamic variation enrichment:} Simulate this by describing rich dynamic variation (gradual swells, rhythmic pulses, fades) that the uniform, unchanging audio does not have.
    \item \textbf{Acoustic environment enrichment:} TTA typically produces dry, close-mic audio. Simulate this by describing a vivid acoustic environment (wide open space, reverberant hall) that the dry audio lacks.
    \item \textbf{Semantic category drift:} Replace one key sound in the caption with a different but plausible-sounding substitute (e.g., ``thunder rumbling'' $\rightarrow$ ``a low, distant explosion'').
    \item \textbf{Count / quantity mismatch:} Change the count in the caption to be clearly different from what the audio contains — either more or fewer (e.g., ``a dog barking'' $\rightarrow$ ``several dogs barking back and forth'').
\end{enumerate}

\noindent\textbf{Critical writing rules:}
\begin{itemize}
    \setlength\itemsep{0em}
    \item Match the style and length of the original caption. Do not write more sentences than the original.
    \item Write like a person casually describing what they hear. Never use comparative or evaluative phrasing like ``sounds like'' or ``resembling''.
    \item NEVER use negative constructions like ``no X'', ``without X'', or ``missing''. Describe what IS there.
    \item Apply only 1–2 modifications to avoid making the caption sound unnatural and artificial.
    \item The modifications must target concrete, audible properties — do not change only abstract descriptions.
    \item The modified caption must remain physically plausible and avoid impossible combinations.
    \item The modified caption MUST be clearly and perceptibly different from the original.
    \item Output only the modified caption with no explanation.
\end{itemize}

\vspace{0.5em}
\noindent\textbf{Original caption:} \{caption\} \\
\textbf{Modified caption:}
\end{tcolorbox}
\caption{Prompt of Attribute Perturbation for the Sound subset, simulating text-to-audio (TTA) generation failures.}
\label{tab:prompt_attr_sound}
\end{table*}

\begin{table*}[t]
\centering
\small
\begin{tcolorbox}[colback=gray!10, colframe=black, boxrule=0.5pt, arc=3pt, width=\textwidth]
\textbf{System Prompt:} \\
You are simulating the failure modes of a text-to-music (TTM) generation model. Given a caption that accurately describes a music piece, rewrite it to describe what a flawed TTM model might actually generate — music that diverges from the original in ways characteristic of real TTM errors. The result is a ``negative caption'' describing the TTM model's erroneous output.

\vspace{0.5em}
\noindent\textbf{Common TTM failure modes to simulate} (apply 1–2 that are most applicable):
\begin{enumerate}
    \setlength\itemsep{0em}
    \item \textbf{Instrument substitution:} TTM sometimes generates the wrong instrument, especially within the same family. Replace one instrument in the caption with a different but acoustically related one (e.g., ``electric guitar'' $\rightarrow$ ``bass guitar'', ``piano'' $\rightarrow$ ``organ'').
    \item \textbf{Genre/style drift:} TTM often confuses adjacent genres. Replace the genre or stylistic descriptor with a closely related but incorrect one (e.g., ``jazz'' $\rightarrow$ ``blues'', ``classical'' $\rightarrow$ ``orchestral pop'', ``R\&B'' $\rightarrow$ ``soul'').
    \item \textbf{Tempo/rhythm mismatch:} TTM struggles with precise tempo control. Change the tempo or rhythmic feel to a clearly different one (e.g., ``fast-paced'' $\rightarrow$ ``mid-tempo'').
    \item \textbf{Mood/energy mismatch:} TTM often generates the wrong emotional character. Replace the mood descriptor with a contrasting but plausible one (e.g., ``energetic and driving'' $\rightarrow$ ``tense and urgent'').
    \item \textbf{Texture enrichment:} TTM tends to generate sparse arrangements. Rewrite the caption to describe a richer, fuller arrangement than the real audio contains.
    \item \textbf{Vocal character mismatch:} If vocals are present, change a key vocal attribute — gender, style, or presence (e.g., ``male vocalist'' $\rightarrow$ ``female vocalist'', ``spoken word verses'' $\rightarrow$ ``sung melody'').
    \item \textbf{Background layer omission:} TTM reliably generates the primary melody but routinely fails to produce subtle background layers. Simulate this by ADDING one such background element to the caption, described as quietly present beneath the main sound, that the real audio does not actually contain (e.g., ``piano and bass with a quiet string pad layered underneath''). The added element must be subtle and secondary.
\end{enumerate}

\noindent\textbf{Critical writing rules:}
\begin{itemize}
    \setlength\itemsep{0em}
    \item Match the style and length of the original caption. Do not write more sentences than the original.
    \item Write like a person casually describing what they hear. Never use comparative or evaluative phrasing.
    \item NEVER use negative constructions like ``no X'', ``without X'', or ``missing''. Describe what IS there.
    \item Apply only 1–2 modifications. Stacking too many changes makes the caption sound unnatural.
    \item The modifications must target concrete, audible properties — do not change only abstract descriptions.
    \item The modified caption must remain physically plausible and perceptibly different from the original.
    \item Output only the modified caption with no explanation.
\end{itemize}

\vspace{0.5em}
\noindent\textbf{Original caption:} \{caption\} \\
\textbf{Modified caption:}
\end{tcolorbox}
\caption{Prompt of Attribute Perturbation for the Music subset, simulating text-to-music (TTM) generation failures.}
\label{tab:prompt_attr_music}
\end{table*}

\begin{table*}[t]
\centering
\small
\begin{tcolorbox}[colback=gray!10, colframe=black, boxrule=0.5pt, arc=3pt, width=\textwidth]
\textbf{System Prompt:} \\
You are an expert with extensive acoustic knowledge. Please describe an audio clip based on the following dimensions, and judge whether the audio matches the provided description. Output \textbf{true} (matched) or \textbf{false} (mismatched) in the consistency dimension, ignoring non-style factors (such as audio quality, naturalness, etc.).

\vspace{0.5em}
\noindent\textbf{Evaluation Dimensions:}
\begin{itemize}
    \setlength\itemsep{0em}
    \item \textbf{Gender:} Voice characteristics associated with different gender identities.
    \item \textbf{Pitch:} The perceived frequency of the sound (e.g., ``female high pitch,'' ``male deep pitch'').
    \item \textbf{Speaking Rate:} How fast or slow the speech is, including specific rhythmic patterns.
    \item \textbf{Volume:} The loudness or softness of the speech (e.g., whispering, normal, shouting).
    \item \textbf{Age:} The speaker's life stage inferred from voice characteristics.
    \item \textbf{Articulation:} Whether the pronunciation is clear and precise or slurred and mumbled.
    \item \textbf{Fluency:} The smoothness of speech, reflecting the presence of hesitations or filler words.
    \item \textbf{Accent:} Distinctive pronunciation reflecting geographical or socioeconomic background.
    \item \textbf{Timbre Texture:} Tonal texture (e.g., sweet, hoarse, deep, bright, warm, nasal, raspy).
    \item \textbf{Emotion:} The feeling expressed while speaking, noting any transitions.
    \item \textbf{Intonation:} Pitch variation patterns conveying attitudinal nuances.
    \item \textbf{Personality:} Overall personality inferred from voice traits.
\end{itemize}

\noindent\textbf{Evaluation Criteria:}
\begin{itemize}
    \setlength\itemsep{0em}
    \item \textbf{true (matched):} The audio meets the primary style dimensions consistent with the description, with no obvious deviations.
    \item \textbf{false (mismatched):} At least one key style feature distinctly conflicts with the description.
\end{itemize}

\noindent\textbf{Notes:}
\begin{itemize}
    \setlength\itemsep{0em}
    \item If there is an obvious discrepancy in a certain dimension (e.g., gender or age conflicting), directly judge as \textbf{false}.
    \item Do not blindly trust the provided description; retain your own understanding of the audio first. 
    \item Pay special attention to the speaker's gender, as it is particularly prone to being the opposite of the description.
    \item Pay extra attention to degree words like ``excited'' or ``intense''. The audio is very likely less intense than described. In such cases, judge as \textbf{false}.
    \item Only evaluate \textbf{style consistency}, ignoring pronunciation accuracy and naturalness.
    \item Features not mentioned in the description imply no restrictions and should not affect the judgment.
\end{itemize}

\vspace{0.5em}
\noindent\textbf{Output Format Requirements (JSON):} \\
\texttt{\{\\}
\texttt{\ \ \ "Gender": "...",\\}
\texttt{\ \ \ "Pitch": "...",\\}
\texttt{\ \ \ "Speaking Rate": "...",\\}
\texttt{\ \ \ ...\\}
\texttt{\ \ \ "Consistency": true/false\\}
\texttt{\}}

\vspace{0.5em}
\noindent\textbf{Description and Audio to be Evaluated:} \\
<Insert audio style description here>
\end{tcolorbox}
\caption{Prompt of Filtering for the Speech subset in Bench construction.}
\label{tab:prompt_filter_speech}
\end{table*}

\begin{table*}[t]
\centering
\small
\begin{tcolorbox}[colback=gray!10, colframe=black, boxrule=0.5pt, arc=3pt, width=\textwidth]
\textbf{System Prompt:} \\
You are an expert with extensive knowledge of music theory and acoustics. Please describe a music piece based on the following dimensions, and judge whether the music matches the provided description. Output \textbf{true} (matched) or \textbf{false} (mismatched) in the consistency dimension, ignoring non-style factors (such as audio quality, recording clarity, etc.).

\vspace{0.5em}
\noindent\textbf{Evaluation Dimensions:}
\begin{itemize}
    \setlength\itemsep{0em}
    \item \textbf{Genre/Style:} The overall style or genre of the music (e.g., classical, jazz, rock, folk, electronic, pop, hip-hop, R\&B, world music). If it crosses over multiple genres, explain each. This is the \textbf{most critical} dimension.
    \item \textbf{Mood/Atmosphere:} The overall emotional tone conveyed (e.g., cheerful, sad, solemn, tense, relaxing, romantic, epic, mysterious, ethereal). Describe the dominant mood.
    \item \textbf{Tempo/Beat:} The rhythmic speed (e.g., soothing, mid-tempo, rapid, or BPM estimate) and time signature (e.g., 4/4, 3/4, 6/8, free rhythm). Specify if there are obvious accelerations/decelerations.
    \item \textbf{Instrumentation:} The primary instruments or timbres used (e.g., piano, guitar, violin, string section, synthesizers, drums). Distinguish between lead and background instruments.
    \item \textbf{Vocal Content:} If vocals are present, describe the singer's gender, vocal style (e.g., bel canto, pop, rap, humming, harmony), lyric language, and emotional expression; if it is purely instrumental, state ``purely instrumental, no vocals.''
    \item \textbf{Tonality/Pitch:} Major (bright, positive) or minor (deep, melancholic) characteristics. Specify if the tonality is ambiguous or if the overall pitch is high or low.
\end{itemize}

\noindent\textbf{Evaluation Criteria:}
\begin{itemize}
    \setlength\itemsep{0em}
    \item \textbf{true (matched):} The music meets the primary style dimensions (genre, mood, instrumentation) consistent with the description, with no obvious deviations.
    \item \textbf{false (mismatched):} At least one key style feature distinctly conflicts with the description, causing the overall auditory perception to deviate.
\end{itemize}

\noindent\textbf{Notes:}
\begin{itemize}
    \setlength\itemsep{0em}
    \item \textbf{Form a judgment based on the music itself first, then compare it with the description} — do not let the wording of the description influence your perception of the musical content.
    \item Do not blindly trust the provided description; it has a high probability of conflicting with the music.
    \item Only evaluate the \textbf{dimensions mentioned in the description}. Dimensions not mentioned imply no restrictions and should not affect the judgment.
    \item \textbf{Genre/Style} and \textbf{Mood/Atmosphere} are the most critical dimensions. If there is an obvious conflict, judge as \textbf{false}.
    \item If the description explicitly requires a certain instrument but the music does not reflect it at all, judge as \textbf{false}.
    \item If the tempo/beat contradicts the description (e.g., described as soothing but presents as rapid), judge as \textbf{false}.
    \item If the description requires a major/cheerful vibe but the music is minor/melancholic, judge as \textbf{false}.
\end{itemize}

\vspace{0.5em}
\noindent\textbf{Output Format Requirements (JSON):} \\
\texttt{\{\\}
\texttt{\ \ \ "Genre/Style": "...",\\}
\texttt{\ \ \ "Mood/Atmosphere": "...",\\}
\texttt{\ \ \ "Tempo/Beat": "...",\\}
\texttt{\ \ \ "Instrumentation": "...",\\}
\texttt{\ \ \ "Vocal Content": "...",\\}
\texttt{\ \ \ "Tonality/Pitch": "...",\\}
\texttt{\ \ \ "Consistency": true/false\\}
\texttt{\}}

\vspace{0.5em}
\noindent\textbf{Description and Music to be Evaluated:} \\
<Insert music style description here>
\end{tcolorbox}\caption{Prompt of Filtering for the Music subset in Bench construction.}
\label{tab:prompt_filter_music}
\end{table*}

\section{Prompt for Decompose Instruction}
\label{sec:prompt_decompose}

In this section, we provide the prompts used to generate and filter rubric questions. Table~\ref{tab:prompt_decompose_instruction} shows the prompt for decomposing instructions into atomic, verifiable questions, and Table~\ref{tab:prompt_rubric_filter} shows the prompt for filtering hallucinated or unsupported questions after rubric generation.

\begin{table*}[htbp]
\centering
\small
\begin{tcolorbox}[colback=gray!10, colframe=black, boxrule=0.5pt, arc=3pt, width=\textwidth]
\textbf{System Prompt:} \\
You are a professional acoustic and audio analysis expert. Your task is to decompose a description of mixed audio (which may contain multiple components like speech, sound effects, and music) into several \textbf{atomic, directly verifiable} yes/no questions for subsequent item-by-item verification against the audio.

The input description may be in Chinese or English; please always output the yes/no questions in \textbf{Chinese}.

\vspace{0.5em}
\noindent\textbf{Decomposition Principles:}
\begin{enumerate}
    \setlength\itemsep{0em}
    \item \textbf{Atomicity:} Each question must involve only one speaker / one dimension / one feature. Do not merge multiple features into the same question.
    \item \textbf{Verifiability:} Each question must be verifiable by directly listening to the audio, avoiding subjective or ambiguous phrasing.
    \item \textbf{Faithfulness:} Only decompose features that are \textbf{explicitly mentioned or strongly implied} in the description. Do not add or infer features out of nowhere.
    \item \textbf{Completeness:} All key features in the description must be covered, including the characteristics of each speaker, speech content, sound effect types, music style, etc.
    \item \textbf{Objectivity:} Use objective, neutral language for the questions, avoiding leading phrasing.
\end{enumerate}

\noindent\textbf{Scope of Dimensions to Cover:} \\
The mixed audio description may involve the following dimensions, all of which must be covered:
\begin{itemize}
    \setlength\itemsep{0em}
    \item \textbf{Audio Components:} What components are included (speech, sound effects, music).
    \item \textbf{Speech Dimensions (decompose separately for each speaker):} Gender (male/female), Pitch (high/medium/low), Age (child/youth/middle-aged/elderly), Timbre (hoarse/clear/deep/gentle, etc.), Emotion (calm/excited/sad/angry, etc.), Speaking Rate (slow/medium/fast), and Speech Content (the specific lines actually spoken).
    \item \textbf{Sound Effect Dimensions:} Type (specific sounds like rain, footsteps, gunshots, flipping pages, etc.), Scene/Environment, and Temporal Sequence (chronological relationship).
    \item \textbf{Music Dimensions:} Style (genre/emotion), Instrumentation (primary instruments), and Music Proportion (foreground/background/intermittent).
    \item \textbf{Mixing Relationships:} Hierarchical Structure (foreground/background relationships) and Dynamic Changes (scene transitions, timing of sound effects, etc.).
\end{itemize}

\noindent\textbf{Question Direction Rules (Important):}
\begin{itemize}
    \setlength\itemsep{0em}
    \item \textbf{All yes/no questions must ensure: answering "yes" = the feature matches the caption description.}
    \item Positive features: Ask directly $\rightarrow$ "yes" means it matches.
    \item Negative/missing features (e.g., "no obvious reverberation"): Must be changed to a negative form (e.g., "Is there \textbf{no} obvious reverberation?") $\rightarrow$ "yes" means it matches.
    \item Do not convert negative descriptions into positive questions, otherwise the yes/no semantics will be reversed.
\end{itemize}

\noindent\textbf{Degree Words \& Speech Content Rules:}
\begin{itemize}
    \setlength\itemsep{0em}
    \item \textbf{Degree Words:} When degree words appear (e.g., "very," "extremely," "slightly," "barely audible"), the degree requirement must be retained in the question.
    \item \textbf{Speech Content:} Each explicitly mentioned line of speech must be a separate question (e.g., Dimension: "Speech Content + Speaker\_A", Question: "Did Speaker\_A say the phrase 'Are you okay?'?"). Do not merge multiple lines into a single question.
\end{itemize}

\vspace{0.5em}
\noindent\textbf{Output Format Requirements (JSON):} \\
\texttt{\{\\}
\texttt{\ \ \ "Question\_List": [\\}
\texttt{\ \ \ \ \ \ \{\\}
\texttt{\ \ \ \ \ \ \ \ \ "Dimension": "<Dimension name, e.g., 'Speaker\_A-Gender' or 'Music Style'>",\\}
\texttt{\ \ \ \ \ \ \ \ \ "Question": "<Question content, ending with a question mark>",\\}
\texttt{\ \ \ \ \ \ \ \ \ "Basis": "<Which sentence or word in the description this question comes from>"\\}
\texttt{\ \ \ \ \ \ \}\\}
\texttt{\ \ \ ]\\}
\texttt{\}}

\vspace{0.5em}
\noindent Do not output any content other than JSON.
\end{tcolorbox}
\caption{Prompt for decomposing instruction into atomic, verifiable yes/no questions for evaluation.}
\label{tab:prompt_decompose_instruction}
\end{table*}

\begin{table*}[t]
\centering
\small
\begin{tcolorbox}[colback=gray!10, colframe=black, boxrule=0.5pt, arc=3pt, width=\textwidth]
\textbf{System Prompt:} \\
You are a strict data quality auditor. Your task is to determine whether each evaluation question is fully grounded in the given caption, or whether it contains hallucinated information that is not mentioned in the caption.

\vspace{0.5em}
\noindent\textbf{Caption:} \\
\{caption\}

\vspace{0.5em}
\noindent\textbf{Questions to Check:} \\
\{questions\_text\}

\vspace{0.5em}
\noindent\textbf{Judgment Criteria:}
\begin{itemize}
    \setlength\itemsep{0em}
    \item \textbf{keep}: The content asked by the question can be fully matched to the caption, without adding extra information.
    \item \textbf{remove}: The question contains concepts, attributes, details, or descriptions that are not mentioned in the caption.
\end{itemize}

\noindent\textbf{Important Notes:}
\begin{itemize}
    \setlength\itemsep{0em}
    \item If the question is only a reasonable paraphrase of the caption content or a conversion into question form, mark it as \texttt{keep}.
    \item If the question introduces new information that is not present in the caption, such as an instrument, emotion, or scene that the caption does not mention, mark it as \texttt{remove}.
    \item Only focus on whether the content of the \texttt{"Question"} field goes beyond the scope of the caption.
\end{itemize}

\vspace{0.5em}
\noindent\textbf{Output Requirements:} \\
Strictly output JSON in the following format, and do not output anything else:

\texttt{\{\\}
\texttt{\ \ \ "results": [\\}
\texttt{\ \ \ \ \ \ \{"id": 0, "judgment": "keep/remove"\},\\}
\texttt{\ \ \ \ \ \ ...\\}
\texttt{\ \ \ ]\\}
\texttt{\}}
\end{tcolorbox}
\caption{Prompt for filtering hallucinated or unsupported questions after rubric generation.}
\label{tab:prompt_rubric_filter}
\end{table*}

\section{Prompts for Judge Evaluation}
\label{sec:prompt_judge}

In this section, we present the two prompts used for judge evaluation. Table~\ref{tab:prompt_decompose_judge} shows the prompt for the Decompose Judge, which evaluates the audio based on decomposed atomic questions. Table~\ref{tab:prompt_holistic_judge} shows the prompt for the Holistic Judge, which evaluates the overall consistency between the audio and the instruction.

\begin{table*}[htbp]
\centering
\small
\begin{tcolorbox}[colback=gray!10, colframe=black, boxrule=0.5pt, arc=3pt, width=\textwidth]
\textbf{System Prompt:} \\
You are a professional audio perception evaluation expert. Your task is to carefully listen to the provided audio clip and answer a given set of yes/no questions one by one.

\vspace{0.5em}
\noindent\textbf{Core Rules:}
\begin{enumerate}
    \setlength\itemsep{0em}
    \item \textbf{Listen Only, Do Not Infer:} Answer solely based on the actually perceivable content in the audio. Do not rely on background knowledge or common-sense inference.
    \item \textbf{Unified Direction:} All questions have been formatted such that "yes = the feature matches the description." Please answer directly according to this alignment.
    \item \textbf{Strictness on Ambiguity:} If a feature is difficult to clearly perceive in the audio, answer \textbf{no}.
\end{enumerate}

\vspace{0.5em}
\noindent\textbf{Output Format Requirements (JSON):} \\
Strictly output a JSON array. The length of the array must perfectly match the number of questions, without containing any other content:

\vspace{0.5em}
\noindent\texttt{[} \\
\texttt{\ \ \ \{"id": 0, "answer": "yes" or "no", "evidence": "<Key evidence you heard, 1 sentence>"\},} \\
\texttt{\ \ \ \{"id": 1, "answer": "yes" or "no", "evidence": "<Key evidence you heard, 1 sentence>"\},} \\
\texttt{\ \ \ ...} \\
\texttt{]}
\end{tcolorbox}
\caption{Prompt for the Decompose Judge to evaluate audio clips against the set of atomic yes/no questions.}
\label{tab:prompt_decompose_judge}
\end{table*}

\begin{table*}[htbp]
\centering
\footnotesize 
\begin{tcolorbox}[colback=gray!10, colframe=black, boxrule=0.5pt, arc=3pt, width=\textwidth]
\linespread{0.9}\selectfont 
\textbf{System Prompt:} \\
You are an expert with extensive acoustic knowledge. Please analyze an audio clip. First, identify all audio components it contains (speech/sound/music, multiple choices allowed). Then, describe the actual performance of the audio based on the corresponding dimensions, and judge whether the audio matches the given instruction description. Output \textbf{true} (matched) or \textbf{false} (mismatched) in the consistency dimension, ignoring non-style factors (such as audio quality, naturalness, etc.).

\vspace{0.2em}
\noindent\textbf{Step 1: Audio Component Identification} \\
Determine all component types \textbf{actually present} in the audio (multiple choices allowed):
\vspace{-0.3em}
\begin{itemize}
    \setlength\itemsep{-0.2em}
    \item \textbf{Speech:} Primarily human voice speaking, reading, or narrating.
    \item \textbf{Sound:} Primarily environmental sounds, nature sounds, mechanical noises, object sounds, etc. (Non-music, non-speech).
    \item \textbf{Music:} Primarily musical creations with melodic/rhythmic structures.
\end{itemize}
\vspace{-0.3em}
The identification result determines the subsequent steps: Single Component (execute Step 2, skip Step 3) or Multiple Components (execute Step 2 + Step 3).

\vspace{0.2em}
\noindent\textbf{Step 2: Dimensional Evaluation} \\
\textbf{Form a judgment based on the audio itself first, then compare it with the instruction.} Evaluate the corresponding dimensions for \textbf{each identified component} separately. \textbf{Field Naming:} For single components, use the dimension name (e.g., "Gender"). For multiple components, use a prefix (e.g., "Speech-Gender") and add an "XX-Proportion" field.

\vspace{-0.3em}
\textbf{[Speech] Evaluation Dimensions:}
\vspace{-0.3em}
\begin{itemize}
    \setlength\itemsep{-0.2em}
    \item \textbf{Gender / Pitch / Age:} Voice identity, frequency (high/low), and inferred life stage.
    \item \textbf{Speaking Rate / Volume:} Speed, rhythmic patterns, and loudness.
    \item \textbf{Articulation / Fluency / Accent:} Clarity, smoothness, and distinctive pronunciation.
    \item \textbf{Timbre Texture / Emotion / Intonation:} Tonal texture, expressed feelings, and pitch patterns.
    \item \textbf{Speech Content:} The actual words spoken.
\end{itemize}

\vspace{-0.3em}
\textbf{[Sound] Evaluation Dimensions:}
\vspace{-0.3em}
\begin{itemize}
    \setlength\itemsep{-0.2em}
    \item \textbf{Sound Type:} Core category (rain, footsteps, etc.). \textbf{Most critical dimension.}
    \item \textbf{Scene/Environment \& Main Elements:} Context and key sound elements.
    \item \textbf{Intensity/Density \& Spatial Sense:} Loudness, thickness, and perception of space.
    \item \textbf{Rhythm/Timing:} Temporal characteristics.
\end{itemize}

\vspace{-0.3em}
\textbf{[Music] Evaluation Dimensions:}
\vspace{-0.3em}
\begin{itemize}
    \setlength\itemsep{-0.2em}
    \item \textbf{Genre/Style \& Mood/Atmosphere:} Overall style and conveyed emotion. \textbf{Most critical.}
    \item \textbf{Tempo/Beat \& Instrumentation:} Speed, time signature, and primary instruments.
    \item \textbf{Vocal Content \& Tonality/Pitch:} Singer's style/language, and major/minor characteristics.
\end{itemize}

\vspace{0.2em}
\noindent\textbf{Step 3: Mixing Relationship Evaluation (For Multiple Components Only)} \\
Describe the \textbf{hierarchical relationship and blending}:
\vspace{-0.3em}
\begin{itemize}
    \setlength\itemsep{-0.2em}
    \item \textbf{Mixing-Hierarchical Structure:} Which is foreground and background.
    \item \textbf{Mixing-Coordination \& Dynamic Changes:} Alignment in emotion/style and temporal changes.
\end{itemize}

\vspace{0.2em}
\noindent\textbf{Evaluation Criteria:}
\vspace{-0.3em}
\begin{itemize}
    \setlength\itemsep{-0.2em}
    \item \textbf{true (matched):} Meets primary features; key dimensions align with no obvious conflicts.
    \item \textbf{false (mismatched):} Missing requested component, key dimension distinctly conflicts, or mixing relationships fundamentally contradict expectations.
\end{itemize}

\vspace{0.2em}
\noindent\textbf{Notes:} Do not blindly trust the description. Verify all components exist. Pay special attention to \textbf{gender} (prone to error), \textbf{degree words}, \textbf{Sound Type}, and \textbf{Music Genre/Mood}. Missing explicitly requested features equals \textbf{false}.

\vspace{0.2em}
\noindent\textbf{Output Format Requirements (JSON):} \\
Strictly output a flat JSON dictionary. Example (Speech+Music):
\texttt{\{"Audio Type": "speech+music", "Speech-Proportion": "...", "Speech-Gender": "...", "Music-Genre/Style": "...", "Mixing-Hierarchical Structure": "...", "Consistency": true\}}

\vspace{0.2em}
\noindent\textbf{Instruction and Audio to be Evaluated:} \\
<Insert instruction description here>
\end{tcolorbox}
\caption{Prompt for the Holistic Judge to comprehensively evaluate audio components, dimensions, and mixing consistency.}
\label{tab:prompt_holistic_judge}
\end{table*}

\end{document}